# (47171) 1999 TC$_{36}$, A TRANSNEPTUNIAN TRIPLE

S. D. Benecchi[a,*], K. S. Noll[a], W. M. Grundy[b], and H. F. Levison[c]

[a] Space Telescope Science Institute, 3700 San Martin Dr., Baltimore, MD 21218
[*] Corresponding Author E-mail address: susank@stsci.edu
[b] Lowell Observatory, 1400 W. Mars Hill Rd., Flagstaff, AZ 86001
[c] Dept. of Space Studies, Southwest Research Institute, 1050 Walnut St. #400, Boulder, CO 80302

Pages: 16
Figures: 8
Tables: 6

**Proposed running head:** A Transneptunian Triple

**Corresponding author:**

Susan D. Benecchi
Space Telescope Science Institute
3700 San Martin Dr.
Baltimore MD 21218
susank@stsci.edu
phone: 410-338-4051
fax: 410-338-5090

## ABSTRACT

We present new analysis of HST images of (47171) 1999 $TC_{36}$ that confirm it as a triple system. Fits to the point-spread function consistently show that the apparent primary is itself composed of two similar-sized components. The two central components, A1 and A2, can be consistently identified in each of nine epochs spread over seven years of time. In each instance the component separation, ranging from 0.023±0.002 to 0.031±0.003 arcsec, is roughly one half of the Hubble Space Telescope's diffraction limit at 606 nm. The orbit of the central pair has a semi-major axis of $a$~867 km with a period of $P$~1.9 days. These orbital parameters yield a system mass that is consistent with $M_{sys}$ = 12.75±0.06 $10^{18}$ kg derived from the orbit of the more distant secondary, component B. The diameters of the three components are $d_{A1}=286^{+45}_{-38}$ km, $d_{A2}=265^{+41}_{-35}$ km and $d_B=139^{+22}_{-18}$ km. The relative sizes of these components are more similar than in any other known multiple in the solar system. Taken together, the diameters and system mass yield a bulk density of $\rho=542^{+317}_{-211}$ kg m$^{-3}$. HST Photometry shows that component B is variable with an amplitude of ≥0.17 ± 0.05 magnitudes. Components A1 and A2 do not show variability larger than 0.08 ± 0.03 magnitudes approximately consistent with the orientation of the mutual orbit plane and tidally-distorted equilibrium shapes. The system has high specific angular momentum of J/J'=0.93, comparable to most of the known Transneptunian binaries.

*Subject headings*: Hubble Space Telescope Observations; Satellites — Composition; Kuiper Belt; Satellites of Asteroids; Photometry

## 1. INTRODUCTION

The discovery of a sizable fraction of binaries among small bodies in the solar system, ~15%, leads to the obvious question of whether higher order systems are also common (e.g. Richardson and Walsh 2006; Noll et al. 2008). Two triple systems have been identified among Near-Earth objects: (153591) 2001 $SN_{263}$ (Nolan et al. 2008) and (136617) 1994 CC (Brozovic et al. 2009) and four triple systems are found among the Main Belt asteroids: (45) Eugenia (Marchis et al. 2007), (87) Sylvia (Marchis et al. 2005), (216) Kleopatra (Marchis et al. 2008), and (3749) Balam (Marchis et al. 2008). Two multiple systems have been identified among Transneptunian objects (TNOs): the two small satellites of the Pluto/Charon binary dwarf planet (Weaver et al. 2006), and two small satellites orbiting the dwarf planet Haumea (Brown et al. 2006).

(47171) 1999 $TC_{36}$ orbits the Sun in the 3:2 resonance with Neptune with a heliocentric semi-major axis of 39.668 AU, an eccentricity of 0.227, and an ecliptic inclination of 8.409°. A secondary was identified in 2002 in images obtained by the Space Telescope Imaging Spectrograph (STIS; Trujillo & Brown, 2002). In the discovery images, the secondary, which we will refer to as component B, was found at a separation of 0.36 arcseconds from the apparent primary and was fainter by 2.21±0.01 magnitudes. Between 2003-2006, (47171) 1999 $TC_{36}$ was observed at seven more epochs with the Advanced Camera for Surveys High Resolution Camera (HRC) leading to a determination of the orbit of the secondary with a period of ~50 days (Margot et al. 2005).

There have been two arguments used to suggest that (47171) 1999 TC$_{36}$ might be a triple system. Stansberry et al. (2006, 2008) found an effective surface area for the unresolved system of $d_{sys}=414.6^{+38.8}_{-38.2}$ km from thermal infrared observations made with the Multiband Imaging Photometer for SIRTF (MIPS) instrument onboard the Spitzer Space Telescope. Using a system mass of $M_{sys} = 14.4 \times 10^{18}$ kg (Margot et al. 2005) they derived a bulk density of $\rho = 500^{+400}_{-200}$ kg m$^{-3}$. For mixtures of water ice and silicate rock, a density this low requires a bulk porosity of 50-75%. Stansberry et al. noted large residuals in the PSF fitting of the primary and speculated that it might itself be a close binary, allowing for a somewhat higher density. Since the Stansberry et al paper, a significant number of similar sized solar system objects (including other TNOs) have been measured to also have low densities (e.g. Spencer et al. 2005; Grundy et al. 2007).

Jacobson & Margot (2007) re-examined the HRC images and found that the apparent primary, component A, appeared to be elongated. With a contour-fitting routine, they found an ellipticity of 0.21±0.05 at a wide range of position angles. Attempts to fit the images with a single point-spread function (PSF) did not adequately reproduce the images of the primary. An F-Test showed with 99.9% confidence that two PSF components are required to properly model the primary. However, detailed binary fits were not presented and the relative positions and size of the components remained undetermined. No orbit solution for the central pair was given.

In this paper we provide unequivocal evidence for the triple nature of (47171) 1999 TC$_{36}$, including a consistent determination of the positions and relative fluxes of all three components using all available Hubble Space Telescope (HST) data. We model the images with a PSF-fitting code and find that the system is composed of two nearly equal-sized components separated by 867±11 km with a third component orbiting the central pair at a distance of 7411±12 km (these are average separations from the orbit fit, not sky plane separations).

In Section 2 we define the terminology we use throughout the paper, in Section 3 we describe the observations, and in Section 4 we describe our data reduction and analysis procedures. In Sections 5 and 6 we present our orbit solution for both components and evaluation of the photometric variability of the system. In Section 7 we present a re-interpretation of system measurements in the new three-body configuration followed by our conclusions in Section 8.

## 2. TERMINOLOGY

In discussing a newly identified component of the (47171) 1999 TC$_{36}$ system we recognize the need for a consistent terminology. Throughout this paper we will refer to the apparent primary as "component A" and the previously identified secondary as "component B". When we discuss the individual components of the apparent primary we will refer to them as "component A1" and "component A2".

## 3. OBSERVATIONS

In this work we present a re-analysis of all available HST images[1] of (47171) 1999 TC$_{36}$. (47171) 1999 TC$_{36}$ was first observed with HST in two separate orbits using the STIS CCD camera in 2001. In each orbit sixteen 120 second images were obtained through the clear filter. The two orbits were separated by 9.62 hours. Beginning in 2003, the HRC was used to obtain images in seven HST visits, altogether spanning slightly more than three years. Each visit (a single HST orbit each) included four un-dithered individual observations, 610 seconds in duration, acquired in the F606W filter. During one of the HRC visits, two images were acquired in an F814W filter for color analysis. The circumstances of all the observations are recorded in Table 1.

---

[1] The data are available from the HST data archive at http://archive.stsci.edu

INSERT Table 1 HERE

We also obtained data of (47171) 1999 TC$_{36}$ over two consecutive nights (UT 2008-09-25,26,27) using the Mt. Bigelow 61-inch in Arizona and three consecutive nights (UT 2008-09-30, 10-01,02) using the CTIO 4-m telescope in Chile. The 61-inch data were collected in V and I filters with the Mont4k camera and the CTIO data were collected in a Sloan r' filter with the Mosaic camera, a 8k x 8k CCD array operated in the 2x2 binned mode with a binned-pixel scale of 0.52 arcsec. A total of 125 exposures, 300-600 seconds in duration, were obtained. A log of observations, including comments about atmospheric conditions and the geometric circumstances during our observations, can be found in Table 1. At Mt. Bigelow both nights were photometric. At CTIO the first two nights were photometric with poor seeing [1.5-1.8 arcsec full-width-half-maximum (FWHM)], while the third night had better seeing (0.8-1.5 arcsec FWHM) with variable cirrus. The seeing at Mt. Bigelow was comparable to the first two nights at CTIO. Photometry was possible for the full five night span using on-chip reference stars calibrated on the photometric nights as described in Section 6.3.

## 4. HST POINT-SPREAD FUNCTION (PSF) FITTING

We processed the HRC and STIS data through the standard HST pipeline (Pavlovsky et al. 2006, Dressel, L. et al. 2007) which performs basic image reduction: it flags static bad pixels, performs A/D conversion[2], subtracts the bias, and dark images, and corrects for flat fielding. It also updates the header with the appropriate photometry keywords. The flat-field-calibrated images, subscripted by *flt*, were the ones we analyzed (for both instruments). We use the individual images instead of the distortion-corrected drizzled image in order to obtain 4 separate measurements for each HST epoch which we can examine both graphically and statistically to evaluate the significance of our result. Our fitting process includes modeling the geometric distortion in the HRC PSF.

Our post-pipeline analysis process for the (47171) 1999 TC$_{36}$ images is the same as was described for HRC images in Benecchi et al. (2009; Grundy et al. 2009), with some additional testing to determine the robustness of our model fits for the central pair, as we describe in section 4.3. We adapted the same fitting software for the STIS images and followed the same basic procedures with a few exceptions as noted.

We performed the PSF-fitting analysis in two steps. First we obtained a best-fit model for component *B*. This component of the system is sufficiently far away (6-13 pixels) from component *A* that we could remove it from the image completely for a cleaner investigation of the central object. Next we modeled the central component as both a single and a binary. We compared the results and found that the binary fits provided a statistically significant improvement over the single object fits. For testing purposes, we also analyzed HRC images of (26308) 1998 SM$_{165}$, a TNO binary with similar brightness, separation and relative magnitude of the two components.

---

[2] A/D conversion takes the observed charge in each pixel in the CCD and converts it to a digital number with the appropriate gain setting. The gain for ACS/HRC is 2.216 e-/DN.

### 4.1. Component B

We began our PSF-fitting by estimating the central positions of component *B* and component *A* (that we consider initially as a single object) by eye (good to about 0.5 pixels). We then sum the object flux in a 1.5 pixel radius around each position to get an initial estimate for the scaling of the PSF. We carried out calculations on sub-images 36-50 pixels on a side (the scale of the HRC is 0.025 arcsec/pixel), centered on component A. The STIS sub-images were 24 pixels on a side (the scale of STIS is 0.05 arcsec/pixel). The sub-images were defined to be large enough to encompass both components of the system and to ensure sufficient sky so that the background could be adequately computed. Bad pixels from cosmic rays and hot pixels within the sub-images were flagged and ignored in further calculations prior to determining the background. For the HRC images we multiplied by a "pixel area map" (PAM)[3] correction image prior to the PSF fitting to correct the photometry for geometric distortions in the images; this correction was not applied to the STIS images. At the location of (47171) 1999 $TC_{36}$ the PAM correction is a factor of 1.127.

The initial estimations for position and flux were further refined by creating a series of scaled and oversampled (by a factor of 4) synthetic PSF images using the Tiny Tim software package developed for analyzing Hubble images (Krist & Hook 2004). For the HRC data, the PSFs were generated at the estimated position for each component (to account for the geometric distortion of the HRC) using a stellar spectral distribution chosen to simulate typical TNO colors [model 57 from the Bruzual-Persson-Gunn-Stryker spectrophotometric atlas (Gunn & Stryker 1983) which has a B-V=0.92, close to the median for TNOs (Hainaut & Delsanti, 2002)]. For the STIS images, where geometric distortion is not an issue, a PSF was generated with the same stellar flux distribution as the HRC, but at a single location, then shifted to the position of the object in the data. We determined each of seven components: average background, position (x1, y1) and flux (f1) of the primary and position (x2, y2) and flux (f2) of the secondary individually by minimizing the $\chi^2$ of the residual between the model and the sub-image while all other values were held fixed.

Next we found values for two additional model parameters that account for small thermally-induced focus changes known as "breathing" and for small motions of the spacecraft known as "jitter". There is some ambiguity between the focus and jitter component values – i.e jitter can be masked by a larger focus value and vice versa. We model the breathing first by adjusting the focus (z4) parameter in the input file for Tiny Tim. It is determined in an iterative fashion, bounding the value on both the positive and negative sides of zero (change of focus units in 0.05 step intervals then 0.001 step intervals as the minimum is determined) and investigating the focus values stepwise to minimize the $\chi^2$ residual. When we fit the focus value, we hold all the other parameters fixed at the best fit position, flux and background values from our initial fit. Including the focus parameter in the model has the effect of improving the $\chi^2$ residual by ~20%. Typical focus values were –0.092 to 0.

Next we modeled both the jitter and the non-negligible angular diameter of the objects by convolving the model PSF with a Gaussian smoothing function (telescope jitter+object diameter), in pixel space. For the component diameters derived by Stansberry et al. (2008), the angular diameters of components A and B are 0.017 and 0.006 arcsec, respectively. For models where components A1 and A2 are considered individually the angular diameters are 0.013 and 0.012 arcsec, respectively. This translates to approximately half a pixel in the HRC and a quarter of a pixel in STIS. We minimize the $\chi^2$ residual to determine the jitter value, where the telescope jitter value can change and the object diameter is fixed, in a step-wise fashion similar to the focus value determination. The jitter value is ~0.11 pixels.

---

[3] http://www.stsci.edu/hst/acs/analysis/PAMS/HRCarea.gif

Next we implement an automated fitting process that uses the *amoeba* (Press et al. 1992) routine, which performs multidimensional minimization of a function containing all our position/flux and background variables using the downhill simplex method, to optimize the fits. The routine does not consider focus or jitter. We ran this automated routine iteratively with the focusing and jitter routines until the $\chi^2$ converged. Typically, 4-5 iterations were required to reach a final PSF model.

From the final model parameters we extracted astrometry and photometry for both components. Photometry was extracted for the HRC data from the best-fit PSF as described in Benecchi et al. (2009). Absolute astrometry was extracted at the fitted positions using IRAF's *xytosky* routine and applying the geometric distortion correction table coefficients for the HRC. It was extracted using IRAF's *xy2rd* routine for the STIS images. The astrometric results for component B, relative to A1, can be found in Table 2.

INSERT Table 2 HERE.

### *4.2. Component A*

In order to better evaluate the central object, component *A*, we subtracted the model of component *B* from the original image and worked only with component A. For this portion of the analysis we used a sub-image box that was 26 pixels on a side for the HRC and 20 pixels on a side for STIS.

To determine if component A is really a binary, we started with the highest resolution data, HRC, and fit component A as a single object for a reference point; we fit only the (x1,y1) position, (f1) flux and background starting with the best fit results from our previous model. We ran the automatic, *amoeba*, fitting process and found a result, as expected, similar to our first (component *A* and *B*) fit. While a formal best fit could be found, the reduced $\chi^2$ ($\chi^2_v$ A) were on the order of 3-5 indicating a poor fit.

Next we tested for a binary fit. We set the position of a secondary component a half pixel away from the single fit location and set the flux of each component to half of the total flux from the single fit and modified the factor added to the jitter to account for the diameter of the central component assuming that the components are equal in size. We leave the focus fixed. We then ran the automatic fitting allowing any of the parameters to change (x1, y1, f1, x2, y2, f2 and background) for $\chi^2$ minimization. We find positional consistency among the results of a single visit, within 0.1 pixels (~0.002 arcseconds). The flux ratios values are the least consistent value in the fit, however the ratio is near 1. The primary effect of the flux ratio being near 1 is that it complicates the orbit fitting due to component ambiguity; we discuss this in Section 5.2. Our fitted positions and flux ratios, listed as component A2 relative to component A1, are provided in Table 3 along with the reduced $\chi^2$ for both the single ($\chi^2_v$ A) and binary ($\chi^2_v$ A1, A2) fits. The absolute astrometric results for the pair component A2 relative to A1 are found in Table 2.

INSERT Table 3 HERE.

We performed the same basic analysis on the STIS images, however, we skipped the step of fitting the inner component as a single since the HRC analysis confirmed the existence of two components and our fit for component B gave a reference point for component A as a single. The pixel scale on STIS is about double that on the HRC and as such there is more scatter in the results, in both pixel space and the residuals. The separation of components A1 and A2 with STIS is slightly less than a pixel (~0.7). Additionally, the STIS data were acquired without moving target tracking so in some cases the PSF trails by 0.5-1 pixels. When plotting the positions of the individual STIS data points we found each visit had points scattered in a region spanning about ~20° (we use the average value of all the points in each visit for the astrometric analysis). However, in each visit, the quadrant of the resulting positions was consistent with the A1-A2 orbit determined with the HRC data. Since we observe precession during the timespan of the HRC data and the STIS astrometric points were further way in time and have larger uncertainties, we did not re-fit the orbit to include the STIS data.

### 4.3. Variations in Initial Conditions

In order to check the robustness of our PSF fits with respect to the initial guesses for component positions, and fluxes, we conducted a series of tests. These include: (1) starting the components at wide (a few pixels) separations, (2) starting the components with extreme flux values (f1~$10^6$, f2~0), (3) fitting for a range of different fixed-jitter values, and (4) re-running the focus determination. None of these changes to the inputs of the model result in $\chi^2$ minimization at significantly different positions (less than 0.01 pixels) or fluxes.

### 4.4. (26308) 1998 SM$_{165}$

We also tested our fitting technique on another binary with similar characteristics to (47171) 1999 TC$_{36}$, also observed with the HRC. (26308) 1998 SM$_{165}$ is in a 2:1 mean motion resonance with Neptune, one of 17 such objects so far to be discovered. A binary companion roughly 1.9 magnitudes fainter was found in images obtained using STIS in late 2001 (Brown and Trujillo, 2002). By combining the measured absolute magnitude of the unresolved pair in the V band, H$_V$ = 6.13 (Romanishin & Tegler 2006), and thermal observations made with *Spitzer*, Spencer et al. (2005) found a diameter of 287 ± 36 km for the primary and 96 ± 12 km for the secondary component. Using the system mass from *HST* observations (Margot et al. 2005) yields a system density of $510^{+290}_{-140}$ kg m$^{-3}$.

The observations of (26308) 1998 SM$_{165}$ that we used to test our PSF fitting were obtained as part of the same observing program as the (47171) 1999 TC$_{36}$ HRC data and are identical in terms of filters and exposure times (obtained 2003-2006, 610 second exposures, F606W filter). Likewise, the time of the HST observations the component separations ($s_{26308}$=0.294±0.001, $s_{47171}$=0.306±0.001 arcsec) and size ratios ($R1\_R2_{26308}$=2.75±0.06, $R1\_R2_{47171}$=2.81±0.03 assuming the same albedos for both objects and their components) are nearly identical making (26308) 1998 SM$_{165}$ an ideal test case. Since the HRC images have the highest resolution of all the observations, these were the ones on which we tested our modeling.

We used the same fitting sequence on images of (26308) 1998 SM$_{165}$ as we did for (47171) 1999 TC$_{36}$. We fit the clearly resolved secondary, subtract it out of the image then fit the central component as either a single or a binary. The binary fit for the central component of this object results in identical positions for the components within 0.04 pixels, or alternatively a flux of nearly zero for the test secondary component.

We show in **Figure 1** a typical sample of the fitting results for the analysis step after component B is subtracted from the image for both (47171) 1999 TC$_{36}$, in two filters, and (26308) 1998 SM$_{165}$. In each triple set of images, the left panel displays the observed data, the middle panel displays the best-fit model PSF and the right panel displays the residuals (observed – model). The data and model are scaled identically and the residuals are scaled to increase the dynamic range. It is clear from the left half of the figure panels A and B that component A of (47171) 1999 $TC_{36}$ is not well fit with a single PSF model. However, it is also clear that if the central component of (47171) 1999 $TC_{36}$ were not binary, it would be well modeled by a single PSF as we see from our model fit results for (26308) 1998 $SM_{165}$ (panel C). The residuals from the two component PSF model in the right half of figure yield very similar residual patterns to those for (26308) 19998 $SM_{165}$. In **Figure 2**, we provide a sample of the fitting results from one exposure in each of the seven HRC visits to (47171) 1999 $TC_{36}$ to demonstrate the repeatability of our results. The orientations of the fields (indicated by the directional in the upper left corner of the panel for each visit) are not identical, but are similar, and it is clear that the two components move with respect to each other between observations.

    INSERT **Figure 1** HERE
    INSERT **Figure 2** HERE

## 5. ORBIT FITTING

We used the component positions determined from our PSF-fitting to determine the mutual orbit of the A1/A2 pair and to re-derive the orbit of component B relative to the A1-A2 barycenter (which we define as half-way between components A1 and A2 using a mass fraction of 0.5, or equal masses). We considered each visit a single astrometric position and averaged the measurements using the scatter in the positions to provide a measure of the uncertainties. We fit the period, semi-major axis, eccentricity, inclination, and the angles: mean longitude at epoch ($\varepsilon$), longitude of ascending node ($\Omega$) and longitude of periapsis ($\varpi$). We run a Monte Carlo simulation on the orbit parameters to derive the uncertainties on each of these parameters and calculate the system mass.

### 5.1. Component B

The orbit of component B (Table 4) was determined from astrometric data that included two epochs of STIS and seven epochs of HRC observations. We found that the seven HRC data points that are closest in time gave a result with the smallest reduced $\chi^2$. The same basic orbit was derived when we added the STIS points farther away in time, however the reduced $\chi^2$ of the orbit increased. Such a result is expected if the system is not Keplerian and there are perturbations within the system that our model does not account for, such as the effects of the non-point mass distribution of the inner pair, which could be approximated by including a quadrupole term ($J_2$) in the primary's gravitational field. We don't have enough observations to actually fit for the masses of all 3 bodies, or even for $J_2$, but we note that spreading the mass of A1 and A2 over a oblate spheroid with the equatorial dimensions of their mutual orbit would produce an effective $J_2R^2$ of 37584 $km^2$, the inclusion of which reduces our chi-squared by more than a factor of two, using Danby (1988) equations 11.15.6 (which give the precession of the secondary's orbit assuming all of the mass and angular momentum reside in the primary). With our results, we can formally rule out the direct and mirror instantaneous Keplerian orbits, however, there are clearly other influences affecting the orbit interpretation. Our values for the orbital period, semi-major axis and system mass ($P$=50.302±0.001 days, $a$=7411±12 km, $M_{sys}$=12.75±0.06 x$10^{18}$ kg) are within the uncertainties of those found by Margot et al. (2005; $P$=50.38±0.5 days, $a$=7640±460 km, $M_{sys}$=13.9±2.5 x$10^{18}$ kg); no additional details for their solution have been published.

### 5.2. Components A1 and A2

The orbit of components A1 and A2 (Table 4) were determined using only the HRC dataset, though the STIS positions are consistent with the orbit we find. We noted earlier that the components have very similar fluxes so we allowed component identification at each visit to be a free parameter in our initial orbit fitting iteration. We used the system mass derived from the orbit of component B to constrain the range of periods and semi-major axes searched. We performed a grid search in which the data were permuted over all 64 meaningful permutations for the 7 ACS visits (swapping identities for all 7 visits at once cannot produce a meaningful new permutation, but swapping a subset of them can, which is why there are 64, not 128). For each of these 64 possible permutations of the data, we used amoeba to fit a circular orbit with initial periods ranging from 1 to 2.35 days in 0.0005 day increments. All solutions having chi-squared values below a threshold of 500 were collected and examined by hand. We then ran fits that permitted the eccentricity to be non-zero, as well as fits for the mirror orbit. The solution appearing in Table 4 emerged as the best one with a $\chi^2$ of 42.6. The next best solution had $\chi^2$ only a little worse, at 89, but a mass nearly double that of the system mass from the orbit of B, so it seemed unlikely to be correct. The next best solution with a similar system mass had a $\chi^2$ over 200, considerably worse that our preferred solution. As we found with the orbit of component B, the smallest $\chi^2$ results from observations taken most closely in time. We find evidence for orbit perturbation with the system over the 3+ years of data included in our analysis. Figure 3 plots the best fit orbits of the two components.

INSERT Table 4 HERE
INSERT **Figure 3** HERE

We calculate the system mass to be $M_{sys}$ = 12.75±0.06 $10^{18}$ kg using the component B system parameters and $M_{sys}$ =14.20±0.05 $10^{18}$ kg using the component A1 and A2 system parameters. The discrepancy in our results are not completely unanticipated, similar findings were reported by Buie et al. (2006) for the Pluto system before the orbit was properly modeled to account for mutual perturbations among the bodies within the system (Tholen et al. 2008). We expect a similar solution exists for (47171) 1999 TC$_{36}$, but such analysis is beyond the scope of this paper. If our solution for the system mass calculated from the component B orbit was correct, and the system was not perturbed, the mass of component *B* would be 0.746±0.001 $10^{18}$ kg.

In addition, based on the orbits we fit, the mutual event season for component B is centered on 2069 and on 2075 for components A1 and A2. However, we have assumed instantaneous Keplerian orbits for this projection excluding precession and we know that mutual perturbations can be expected to shift the timing somewhat, perhaps on the order of a decade.

## 6. PHOTOMETRY

### 6.1. Previous photometric observations

(47171) 1999 TC$_{36}$ is a relatively bright TNO and therefore has been the target of a number of photometric studies. Attempts to determine the lightcurve of the (47171) 1999 TC$_{36}$ system (combined light of all components) have found it to be inconsistent across epochs. Peixinho et al., 2002, observed (47171) 1999 TC$_{36}$ for 8.362 hrs and found no apparent periodic variation. Ortiz et al., 2003, found variation with a period of 6.21±0.02 hours, but were unable to obtain repeatable observations a month later. Its phase curve changes with wavelength from

B'=0.255±0.044 mag/° to V'=0.131±0.049 mag/° to I'=0.235±0.056 mag/° (Rabinowitz et al. 2007). Barkume et al. (2008) and Dotto et al. 2003 found the surface to have a moderate ~8-20% ice fraction signature when observed spectroscopically. The visible colors of the unresolved A1+A2 pair, and component B are nearly identical and red, $(V-I)_A=1.19±0.01$ and $(V-I)_B=1.12±0.03$ (Jacobson, private communication 2007; Benecchi et al. 2009).

### *6.2. HST variability*

HST observations are intrinsically photometric. The sparse sampling obtained by HST, while unsuitable for lightcurve measurement, can be used to search for and constrain possible variability. In **Figure 4** we plot the photometry of the largest dataset in a single filter, HRC F606W. The data cover 3 years of time. The measurement uncertainties are ~0.01 magnitudes and the variation within each 30 minute visit is small (0.01-0.05 magnitudes). When all the data are considered, the variation of component B is 0.17±0.05 magnitudes. The variation of component A1+A2 is 0.08±0.03 magnitudes. We suggest that if the components are resolved, and measured on a shorter timescale, both will show lightcurves with amplitudes greater than the variability found in the HST dataset. Because of the dynamics in the system, we hypothesize that component A1 and A2 are tidally locked with rotational periods of 45.74 hours.

INSERT **Figure 4** HERE

### *6.3. Ground based photometry*

We obtained two nights of ground-based data of (47171) 1999 $TC_{36}$ at the Mt. Bigelow 61-inch telescope and three nights at the CTIO 4-m Blanco telescope (Table 1) in September/October 2008. The data were bias-subtracted and flat-fielded with standard procedures. We performed aperture photometry on (47171) 1999 $TC_{36}$ as a system as well as about a dozen field stars for comparison. (47171) 1999 $TC_{36}$ did not come close to any other stars during the five nights of our observations. The seeing throughout the runs was 1.0-1.8 arcseconds (2-4 pixels) and as such we used a photometric aperture radius of 12-14 pixels. We found that the individual measurements of (47171) 1999 $TC_{36}$ yield S/N~90 for the CTIO data and S/N~10 for the 61-inch data (which we binned by 3 for analysis), and performed aperture correction photometry on our object to reduce background contamination, using the two most photometrically stable comparison field stars (at the level of 0.01 magnitudes). For the 61-inch observations we photometrically calibrated our object and the field stars with the Landolt standard star, SA 113_260 (Landolt, 1992). We performed similar analysis on nights one and two of the CTIO observations using the Landolt standard star, SA 113_339 (Landolt, 1992). We used the field stars to photometrically calibrate the data from the nights of our runs that were not photometric (night one at the 61-inch and night three at CTIO). Lastly, we shifted the 61-inch data which was collected in V and I filters to the Sloan r' filter using a combination of measured offsets from the data and *synphot* modeling so that all the data could be analyzed together.

In **Figure 5** we plot the lightcurve from each night of data in a separate panel referenced to the first observation. The variation of the data is 0.20±0.04 magnitudes. We attempted to model the data with a weighted single peaked lightcurve with periods ranging from T=3-24 hours (6-48 hour double peaked periods), but do not find any satisfactory results. Periods of ~16±2 and 23±2 hours are the most prominent, but the residuals are not clean and the 23 hour period, which is about what we expect for a tidally locked system, is too close to our observing interval of 24 hours to be confident of its validity. The uncertainty in the amplitude of the model is ~30% of the amplitude itself, which ranges between 0.03±0.01 and 0.06±0.03 magnitudes for our two best-fit periods. It may be the case that the tidally locked period is correct for components A1 and A2, and a contribution of variation from component B complicates our ability to find a clean fit, we cannot say for certain.

From Leone et al. 1984 we find that the equilibrium tidally distorted shape for two components of equal size with $\omega^2/\pi G\rho = 0.012$ (where ω is 2π/T and ρ is density) will have $a/c \sim 0.97$ (where $a$ and $c$ are the smallest and largest diameters of the triaxial ellipsoid). The maximum possible lightcurve amplitude for an object of this shape is 0.03 magnitudes. However, because the orbit is inclined, the amplitude is reduced to 0.02 magnitudes by the cosine of the angle of the normal to the line of sight, 40°. Since components A1 and A2 dominate the photometric signature, the lack of a strong lightcurve is therefore not surprising.

Limits for the effect of the variability of component B on the unresolved system brightness can be calculated. The average delta between components A1+A2 and component B is 2.24±0.03 magnitudes (Table 6) and the average brightness of the system is 19.50±0.01 magnitudes. If component B varies by as much as 2 times that found from the HST images, it would contribute a maximum 0.04 magnitude amplitude to the unresolved system lightcurve ($M_{TOTAL}$ would range between 19.50 and 19.54). The variability we find for our combined system lightcurve are consistent with this range of variability for component B, though the period and contribution of each component to the variation remains unknown. Components A1 and A2 cannot have large (>0.2 magnitude) variations as we would have measured them in our data.

INSERT **Figure 5** HERE

## 7. DISCUSSION

### 7.1. *Comparison to other multiple systems*

In the context of the other known multiple systems in the solar system's small body populations, (47171) 1999 TC$_{36}$ demarks an extreme in relative component sizes. It is composed of two nearly identically sized components in close proximity with a third component approximately half the size of components A1 and A2 nearly 10 times farther away. All the other known small body multiples contain at least one relatively tiny component (Table 5, **Figure 6**) and components are approximately evenly spaced. In most systems the more distant of the two outer components is the larger, which we note is not true for (47171) 1999 TC$_{36.}$

INSERT Table 5 HERE
INSERT **Figure 6** HERE

In absolute terms, the known multiple systems range over almost three orders of magnitude in radius and over eight orders of magnitude in volume. Despite this very large range of absolute sizes, the system architectures are remarkably similar (in the appropriately scaled units). All of the systems are tightly bound, with components orbiting at a few percent of the Hill radius or less. With only 9 known multiples in the solar system, it is too early to know whether there are clear groupings of systems in terms of their physical and/or orbital properties or, instead, if there is a continuum of such systems. If the former, the architecture of multiples may hold clues to their modes of formation.

### 7.2. *Density and Porosity*

The confirmation of (47171) 1999 TC$_{36}$ as a triple impacts the derivation of its physical properties, in particular its density and porosity. These issues were first discussed by Stansberry et al. (2006). We revisit these calculations using the newly determined relative brightnesses of components A1 and A2 to remove this formerly speculative quantity.

To calculate the density and porosity of (47171) 1999 $TC_{36}$ we used an effective system diameter of $d_{sys}=414.6^{+38.8}_{-38.2}$ km (Stansberry et al. 2008). We assume that the components have equal albedos, 7% (Stansberry et al. 2008), and determine their individual diameters from their observed flux differences which are found in Table 6. We average all the F606W measurements to find $\Delta m_{AB}$=2.24. From this we determine the diameter of component B to be $d_B=139^{+22}_{-18}$ km. Similarly, we find $\Delta m_{A1A2}$=0.17 yielding the individual diameters of the two inner components, $d_{A1}=286^{+45}_{-38}$ km and $d_{A2}=265^{+41}_{-35}$ km. The similarity of the sizes of the components is, so far, unique among the multiple systems in the Kuiper Belt.

INSERT Table 6 HERE

We calculate the density of the system as:

$$\rho = \frac{3M}{4\pi\left[(d_{A1}/2)^3 + (d_{A2}/2)^3 + (d_B/2)^3\right]}$$

(1)

and find $\rho=542^{+317}_{-211}$ kg m$^{-3}$.

For a range of material densities $1000 < \rho_0 < 2000$ kg m$^{-3}$, we find porosities, or fractional void space $f = 1 - \rho/\rho_0$, of 46-73%. (47171) 1999 $TC_{36}$, however, is not the only TNO with a low density. (26308) 1998 $SM_{165}$ has a density of $700^{+320}_{-210}$ kg m$^{-3}$ (Spencer et al., 2006), and (42355) Typhon has a density of $440^{+440}_{-170}$ kg m$^{-3}$ (Grundy et al. 2008). Low densities have also been reported for some of the icy satellites (Burns, 1986; McKinnon et al., 1995) and the Trojan asteroid (617) Patroclus (Marchis et al. 2006; Mueller et al. 2009).

### 7.3. Angular Momentum

The specific angular momentum of the (47171) 1999 $TC_{36}$ system can be calculated by summing the orbital and spin angular momenta of the components and normalizing by the factor $J' = \sqrt{GM_{sys}^3 R_{eff}}$ where G is the gravitational constant, $M_{sys}$ is the system mass and $R_{eff}$ is the radius of an equivalent spherical object containing the total system mass. The total angular momentum of (47171) 1999 $TC_{36}$ is dominated by the orbit angular momentum given by $J_{orbit} = \sum_i m_i a_i^2 \omega_i \sqrt{(1-e_i^2)}$, where $m_i$ is the mass of each component, $a_i$ is measured from the system barycenter, $e_i$ is the eccentricity of the orbit, and $\omega_i$ is the orbital frequency. For this calculation we assumed that all of the components have the same density. The spin component of the angular momentum is given by $J_{spin} = \frac{2}{5}\sum_i m_i r_i^2 n_i$ where $n_i$ is the rotational frequency of each component. For this calculation we assumed A1 and A2 to be tidally locked so that their spin period is equal to their 1.906 day mutual orbit period. The more distant component B's spin period is unknown; we assumed a period of 8 hours, typical for small bodies. With these assumptions, the spin component contributes only 2.6% of the total angular momentum. In the extreme case all of the components are assumed to be rotating near break-up with a period of 3 hours; the spin component of the angular momentum rises to 27.7% of the total. All of the possible cases are qualitatively the same: the system's total angular momentum is dominated by the orbit angular momentum. Additionally, the independent contribution to orbital angular momentum is similar, though not identical, for each of the three components. Component A1 contributes 33.7%, component A2, 26.4% and component B, 39.9%.

Total system angular momentum can provide clues to the mode of formation of multiple systems (**Figure 7**). Systems formed by spin-up and fission must have J/J'<0.39 (Dobrovolskis et al. 1997). Many near earth and main belt systems may have formed this way. Multiples formed by collision have an upper limit of J/J' ~0.8 (Canup 2005). This assumes that ratio of the impactor velocity to the escape velocity is $v_{imp}/v_{esc} < 1.3$. While this is probably a safe assumption for large systems thought to have formed by collision like Pluto (Canup 2005) and Haumea (Levison et al. 2008), it is less clear whether this limit applies to smaller systems like (47171) 1999 TC$_{36}$ where the escape velocity, ~0.1 km/s, is likely lower than random velocities in the disk. Most Transneptunian binaries (TNBs) have J/J' >0.8. Formation by dynamical capture (Goldreich et al. 2002, Noll et al. 2008) allows for higher system angular momentum and thus may be more consistent with the high angular momenta found in TNBs. Like most TNBs (47171) 1999 TC$_{36}$ has a relatively large angular momentum, J/J'=0.93, and we suggest that it likewise may have been formed by two successive dynamical capture events.

INSERT **Figure 7** HERE

### 7.4. Non-Kelperian Orbits

A full n-body solution to the orbital dynamics of the (47171) 1999 TC$_{36}$ system requires more astrometric data than is currently available and is therefore beyond the scope of this paper. However, it is possible to investigate the general dynamical behavior of this system using our estimates of system parameters as a starting point for a model of the orbital evolution. As shown in **Figure 8** we found that $\varpi$ precesses through a full $2\pi$ radians in roughly 50 years.

INSERT **Figure 8** HERE

### 7.5. Tidal Evolution

For a two-component system, the time to circularize an initially eccentric orbit can estimated from (Goldreich & Soter 1966, eq. 25):

$$T_{circ} = \frac{4Qm_2}{63m_1} \sqrt{\frac{a^3}{G(m_1+m_2)}} \left(\frac{a}{r_2}\right)^5.$$

(2)

The semi-major axis of the orbit, $a$, the radius of the smaller component, $r_2$, and the component masses, $m_1$ and $m_2$, are available from fits to the mutual orbits (Table 4 and derived values) and the tidal dissipation coefficient, Q, is estimated to range from 10-500 for rocky and icy bodies (Goldreich & Soter 1966). However, as noted by Noll et al. (2008) this equation has limited applicability for systems with large secondaries. To apply this formulation to (47171) 1999 TC$_{36}$ we ignore component B when considering the evolution of the A1 and A2 orbit or treat A1+A2 as a point mass when consider the tidal evolution of component B. To the extent that these approximations violate the assumptions leading to equation 2, these results should be used with caution.

The circularization timescale for component B is $T_{circ} \sim 10^8$ years. We note that with an eccentricity of 0.2949 the system is not circularized, although given the large uncertainties in the tidal circularization formulation, this may not be constraining. We find an extremely short circularization timescale of $T_{circ} \sim 100$ years for the inner pair (A1 and A2). Even allowing for order of magnitude uncertainties, the inner pair would be expected to have evolved to zero eccentricity in the absence of other perturbers. The non-zero eccentricity of the A1 and A2 pair in our orbit solution points to complex dynamical interactions between the three components.

### *7.6.    Formation*

The formation of multiples can be considered a special case of the more general problem of how to form binaries. For the Kuiper Belt several possible modes of formation have been proposed including collision (Canup 2005) and gravitational capture (Goldreich et al. 2002). We consider each in the following sections. Additionally, Nesvorny (2008) has suggested that TNBs may have formed directly during the gravitational collapse of the gaseous protoplanetary disk when the excess of angular momentum prevented the agglomeration of all available mass into solitary objects. The details of this model are not yet available.

#### *7.6.1.    Collision*

Two other multiple systems are known in the Kuiper Belt. The Pluto system consists of two large binary components, Pluto and Charon, with two more small satellites, Nix and Hydra. Nix and Hyrda orbit near the 4:1 and 6:1 mean motion resonances (formally 3.991±0.007:1 and 6.064±0.006:1), respectively and share an orbit plane with Charon (Tholen et al. 2008). These details are consistent with formation by a giant collision and reaccretion in a disk (Ward & Canup, 2006; Lithwick & Wu, 2008). It has to be noted, however, that this consistency does not constitute proof that this system arose from a collision (e.g. Goldreich et al. 2002).

Haumea and its two small satellites Hi'iaka and Namaka are members of a family of collision fragments identified by their strong water ice absorption features (Brown et al. 2007; Ragozzine & Brown 2007). Dynamical evolution models find that such a family could not have been created in the massive primordial belt because the coherence of the family members would not have survived to current times. Levison et al. (2008) explored the formation of a family resulting from the collision of two scattered disk objects and found the existence of one such family consistent with their results. It seems nearly certain that this triple system arose from a low probability collision (Levison et al. 2008; Schlichting & Sari 2009).

The (47171) 1999 $TC_{36}$ triple system differs significantly in its architecture compared to Pluto and Haumea (**Figure 6**). The components are more similarly sized and there is an order of magnitude difference in their semi-major axes. The angular momentum of the (47171) 1999 $TC_{36}$ system is substantially higher than for the Pluto and Haumea systems (**Figure 7**) and appears to be more comparable to the larger population of TNBs, as discussed in Section 7.3.

#### *7.6.2.    Gravitational capture*

Gravitational capture by multi-body events in a dense planetesimal disk (Goldreich et al. 2002) produces binaries with high angular momenta and with an increasing fraction at small separations as is observed (Noll et al. 2008). Equal-sized components are preferred in at least some capture models (Astakhov et al.2005, Lee et al. 2007), again in accord with observations. Based on these agreements between model and observation, we have suggested that most TNBs are formed by capture (Noll et al. 2008).

Goldreich et al. (2002) specifically address the formation of multiples and predict that such systems should exist as a natural result of hardening caused by dynamical friction and ongoing capture events. For (47171) 1999 $TC_{36}$ the picture is an initial binary system in the process of hardening, but securely captured, encountering another pair of objects (or a single similarly sized object in a sea of small bodies) resulting in a second component being captured and likewise hardening.

Successive capture events will initially have an orbit plane oriented randomly to a preexisting pair. However, it is possible that initially different orbit planes will evolve. For example, a system that starts off with near-orthogonal orbit planes would experience strong Kozai resonances that would result in episodes of high eccentricity. Such interaction could lead to tidal damping, coalescence or perhaps loss of the component. If non-coplanar orbits are more easily disrupted, the existence of a nearly coplanar triple like (47171) 1999 TC$_{36}$ may simply be a selection effect.

### *7.7.     Are there more triples and how do you find them?*

Finally, we ask, how do we find additional multiples in the Kuiper Belt, assuming that such systems exist? It has been previously shown (Kern & Elliot 2006; Noll et al. 2008) that the separation distribution for TNBs increases with decreasing separation to the observation limit (0.05 arcseconds). At the current observation limits, direct imaging of close triples is unlikely if they are composed of three approximately equal sized components with two components close to contact, as is the case for (47171) 1999 TC$_{36}$.

Lightcurves of known binaries have the potential to reveal an additional unresolved component. This technique relies on the assumption that an unresolved pair will be tidally locked and have a lightcurve period equal to the orbit period. Typically, this will be longer than the range of rotational periods for small bodies (4-18 hrs, Lacerda & Luu, 2006). While such a method does not result in an unambiguous identification of multiples, it can identify candidates for higher resolution follow-up and provide statistical information.

Ideally, a search for multiples would result from resolved lightcurves of both primary and secondary components using HST. Unresolved observations can also be used to identify multiples by fitting a complex lightcurve. However, the added complication of accounting for two variable components with different periods increases the demands on photometric precision.

Indirect detection of binaries through lightcurve analysis was pioneered by Pravec et al. (2002) and is now the most productive method for identifying binaries in the Near Earth asteroid population (Pravec et al. 2006). In the transneptunian population, Sheppard & Jewitt (2004) indirectly identified a possible close binary from its lightcurve and from it estimated that ~15% of TNOs may be similarly close pairs.

Alternatively, determination of a binary orbit with non-Keplerian motion may indicate an additional component in the system. While no systems have been identified by this method, both the Haumea (Ragozzine & Brown, 2009) and (47171) 1999 TC$_{36}$ triple systems have non-Keplerian orbits.

## 8.     CONCLUSIONS

We present new compelling evidence that (47171) 1999 TC$_{36}$ is a triple system. The component originally identified as the primary consists of two components that orbit each other with a period of T$_{A1A2}$ = 1.9068±0.0001 days at a separation of a$_{A1A2}$ = 867±11 km. Components A1 and A2 have diameters of of $d_{A1}=286^{+45}_{-38}$ km, $d_{A2}=265^{+41}_{-35}$ km assuming all components share the same albedo of 7%, determined from Spitzer measurements (Stansberry et al. 2008).

We have also re-determined the orbit of the more distant secondary, component B. We find an orbital period of T$_B$ = 50.302±0.001 days and a semimajor axis of a$_B$ = 7411±12 km in approximate agreement with previous work. The diameter of component B is $d_B=139^{+22}_{-18}$ km for the same assumed albedo. From the component B orbit we find a system mass of $M_{sys}=$ 12.75±0.06 $10^{18}$ kg.

An orbital analysis of this system demonstrates that it precesses rapidly with $\varpi$ completing a full $2\pi$ in approximately 50 years for all of the components. This rapid precession limits our ability to find a unique Keplerian orbit solution (valid on the timescale of years) based on the current data.

Using the derived component diameters we find a system density of $\rho=542^{+317}_{-211}$ kg m$^{-3}$. A porosity of 46-73% is required for a range of material densities from 1000-2000 g/cm$^3$. The relatively large uncertainty in the density is primarily due to the uncertainties in the measurements of the thermal emission by Spitzer (Stansberry et al. 2008). The system has a high specific angular momentum, J/J'=0.93 and is likely to have been formed through gravitational capture.

The (47171) 1999 TC$_{36}$ system does not demonstrate a simple lightcurve over five nights of ground-based observations. The photometric data shows variability of 0.20 ± 0.04 magnitudes for the unresolved system. The HST photometry shows that component B varies by at least 0.17 ± 0.05 magnitudes and components A1 and A2 combined vary by at least 0.08 ± 0.03 magnitudes. Attempts to model the lightcurve of the unresolved system do not yield statistically significant results.

## ACKNOWLEDGMENTS

We thank M. Buie for the use of his PSF fitting code in IDL (Grundy et al. 2009) and his advice on how to go about modifying it for the HRC images. We also thank Dr. Don McCarthy and students Nancy Thomas and Shae Hart for their contributions of the 61-inch data and Astronomy Camp for supporting their work. This work is based on observations made with the NASA/ESA Hubble Space Telescope. These observations are associated with program #9746. Analyses were supported by HST programs #11178 (PI. Grundy) and #11113 (PI. Noll). Support for these programs was provided by NASA through a grant from the Space Telescope Science Institute, which is operated by the Association of Universities for Research in Astronomy, Inc., under NASA contract NAS 5-26555.

**FIGURES**

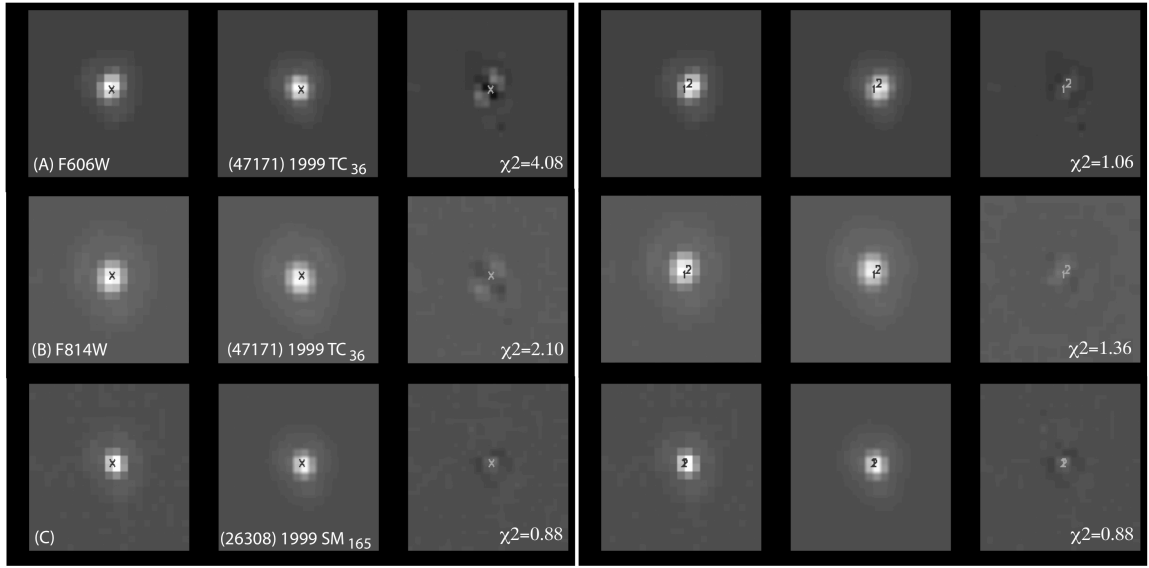

**Figure 1. Sample fitting results for (47171) 1999 TC$_{36}$ in both F606W (panel A) and F814W (panel B) filters and (26308) 1998 SM$_{165}$ (panel C).** Each triplet of images from left to right includes: observed data, PSF model and residual image (observed-model) with the left row showing the results of fitting the object with a single component ('X' marks the best fit location) and the right fitting it with two components. Numbers (1, 2) are plotted at the pixel location where the position and flux of the components minimize the $\chi^2$ residual. The $\chi^2$ residual for the binary fit of (47171) 1999 TC$_{36}$ is a factor of 2-4 times better than that of the single fit, independent of filter. For (26308) 1998 SM$_{165}$ the binary fit results in two components with identical locations with a nearly identical $\chi^2$ to the single fit. The image range is 13000 counts.

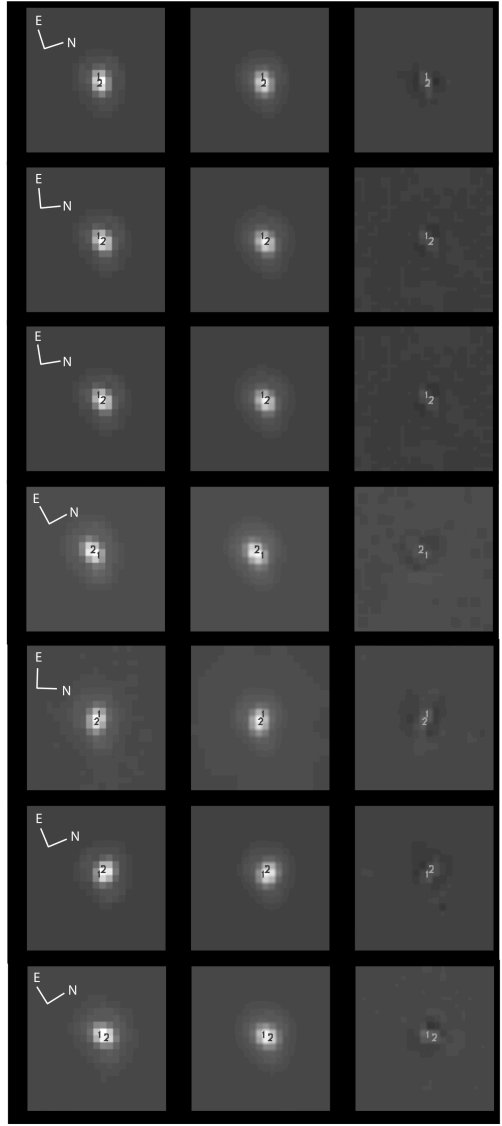

**Figure 2. Sample results of binary fits.** The images are ordered by time. From left to right, the panels are: observed data, Tiny Tim model and residual image. Numbers (1, 2) are plotted at the pixel location where the position and flux of the component minimize the $\chi^2$ residual. The orientation of all the images is similar, but not identical, so a North-East indicator has been added for each visit. In most cases, 1 corresponds to the brighter of the two components.

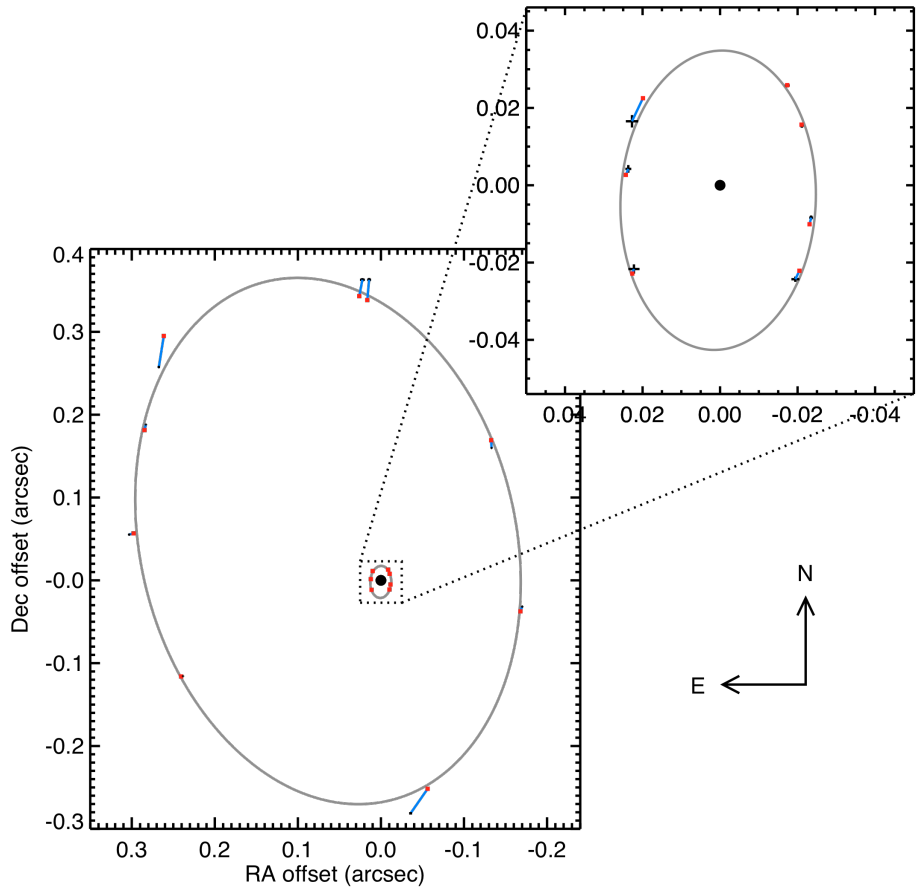

**Figure 3. Component orbits.** Component B has a period of 50.302±0.001 days and component A2 has a period of 1.9068±0.0001 days (shown to scale and blown up for visibility). The panel with both orbits (lower left) is plotted relative to the A1-A2 barycenter, while the zoomed panel (upper right) is plotted relative to A1 (this conserves the correct scaling of the orbits and uncertainties). Each black point with error bars (often difficult to discern) is an average of the points collected during a single HST orbit. The model is a weighted fit to the points, and the blue lines drawn between the black data points and the red model points indicate the residuals at the time of observation. A filled black circle represents A1 in the upper right panel and the A1-A2 barycenter in the lower left panel. North is up, and East is to the left.

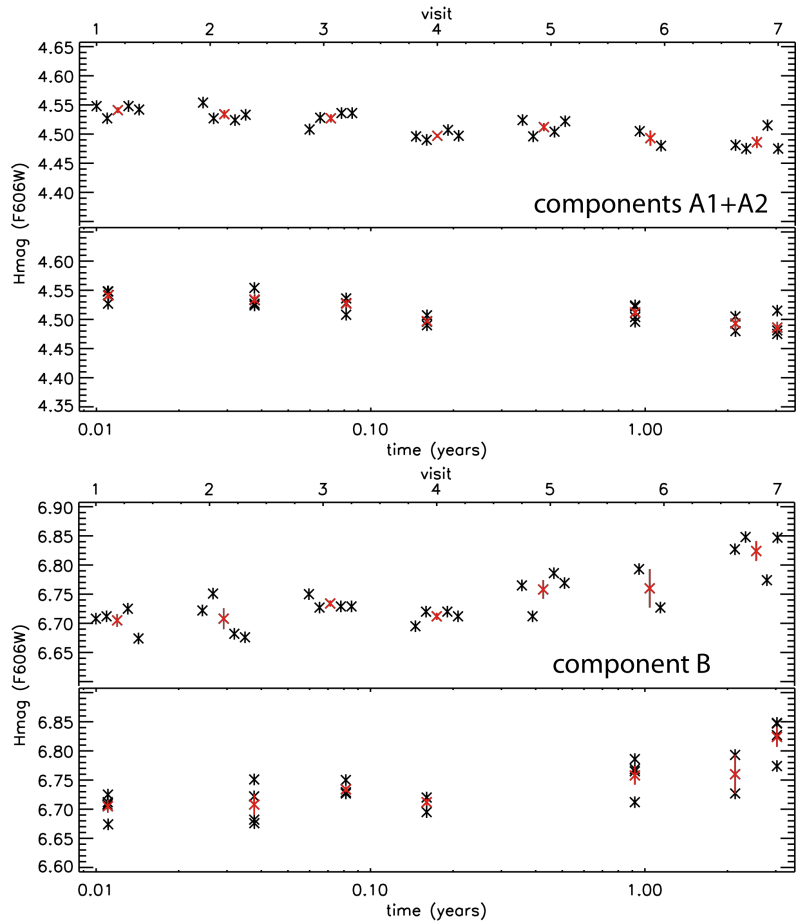

**Figure 4. Resolved variability for the inner pair (top) and component B (bottom).** The photometry is geometrically corrected and plotted by HST visit, on a log scale to facilitate better visibility. Time zero is defined as the time of the first HST observation (offset by 0.01 days since log 0 is undefined) and average values for each visit are plotted in red. The top panel of each plot zooms in on each visit to demonstrate the scatter in the points, which are used within each visit to calculate the uncertainty on the average values. The unresolved component A1+A2 has a peak-to-peak variation of 0.08±0.03 magnitudes and component B has a peak-to-peak amplitude of 0.17±0.05 magnitudes.

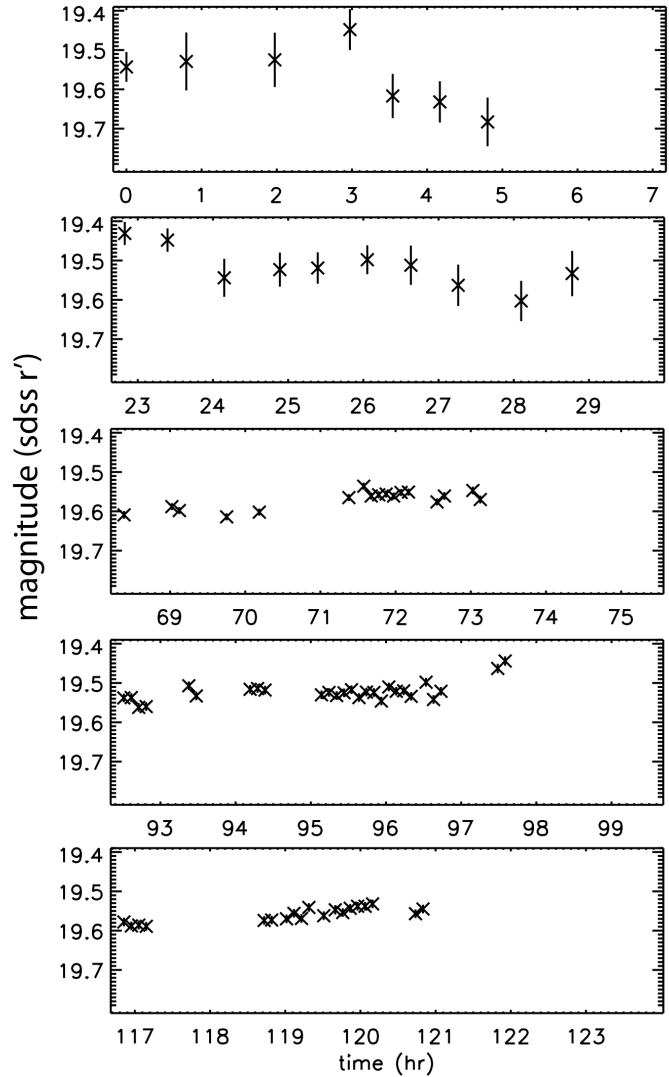

**Figure 5. Ground-based lightcurve observations of (47171) 1999 TC$_{36}$.** Data were collected from the Mt. Bigelow 61-inch and CTIO 4-m telescopes over 5 nights. The data have been plotted so that $t_0$ is the time of the first observation. The 61-inch data were obtained using the Mont4k camera and the CTIO data were obtained using the Mosaic camera. The 61-inch data are normalized to the Sloan r' filter and the uncertainties in the CTIO data are about the size of the data points. Attempts to model the data with a lightcurve do not yield any statistically significant results. The 61-inch data were contributed by Dr. Don McCarthy and students Nancy Thomas, Shae Hart of the University of Arizona Astronomy Camp.

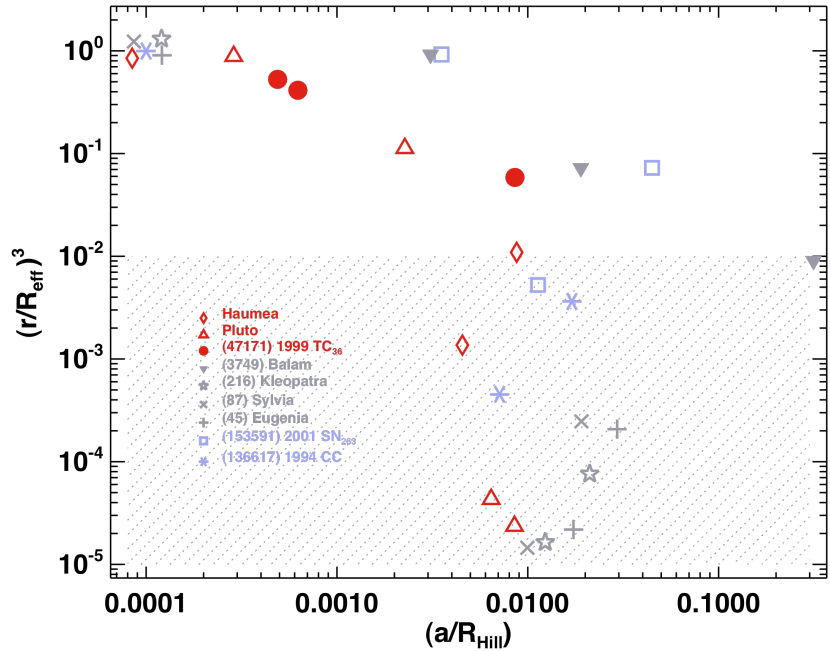

**Figure 6. Comparison of Solar System multiples.** All eight known multiples (n>2) in the solar system's small body populations are represented in this plot showing the relative sizes and separations of the components. The semimajor axis relative to the system barycenter, $a$, scaled to the Hill radius, $R_{Hill}$, is plotted on the x-axis with a logarithmic scale. The relative component mass is plotted on the y-axis, also with a logarithmic scale, as the cube of the ratio of the component radius, $r$, to the cube of the effective system radius, $R_{eff}$ where $R_{eff}$ is the radius of a sphere having the same volume as all of the components. Primaries were forced to have a minimum $a/R_{Hill}$ of $10^{-4}$ and offset for clarity. Components falling in the stippled area have less than 1% of the volume (and mass) of a body having the effective system radius. (47171) 1999 TC$_{36}$ stands out from the other TNO triple systems in that all three components contribute a significant fraction of the total system mass.

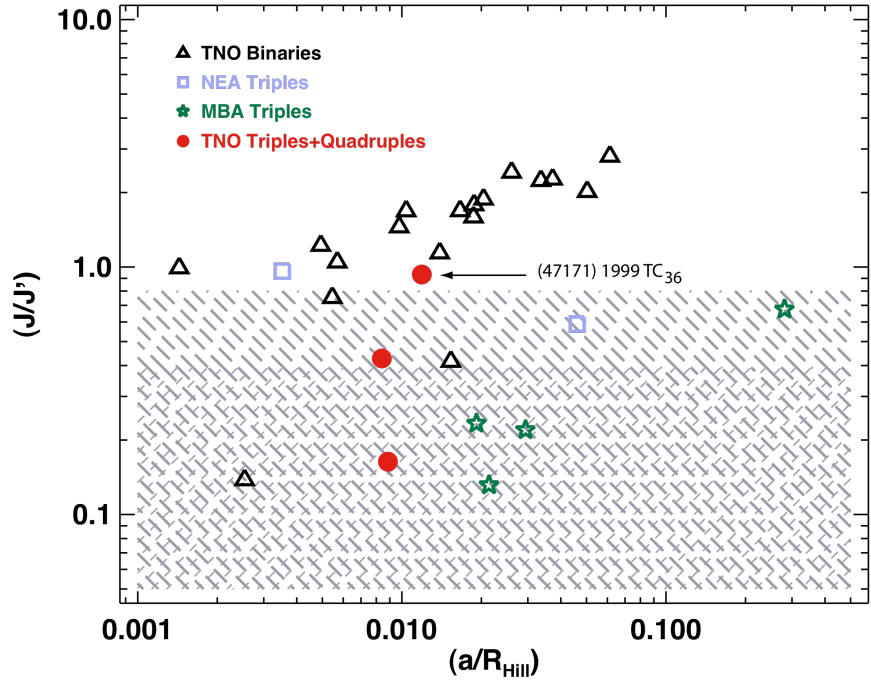

**Figure 7. Comparison of angular momenta for Solar System multiples.** Normalized angular momentum plotted vs. semi-major axis scaled by the Hill radius, $R_{Hill}$, of the system for all binaries and triples in the Kuiper Belt with determined orbits (Noll et al. 2008; Grundy et al. 2009) and all other solar system triple systems. Where rotational periods are known, we use them, where they are unknown we assume a period of 8 hours. Dobrovolskis et al. 1997 find that objects with J/J' > 0.39 cannot have formed through rotational fission. Canup (2005) suggest that such systems are likely to be formed by a single catastrophic collision, however, if J/J' > 0.8 additional considerations exist and formation by collision is limited. Systems with J/J'>0.8, including (47171) 1999 TC$_{36}$, were most likely to formed through three-body gravitational interactions/capture as suggested by Goldreich et al (2002) (Chiang et al. 2007) or via the Nesvorny (2008) formalism.

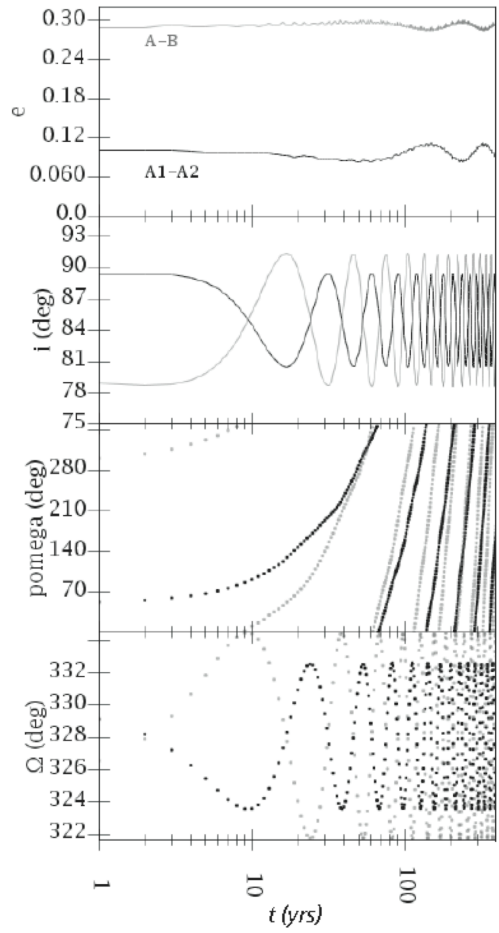

**Figure 8. Orbit evolution.** The evolution of our nominal orbit (Table 4) for both components is displayed. The orbit is stable and $\varpi$ precesses through a full $2\pi$ radians in roughly 50 years.

TABLE 1. OBSERVATIONS

| UT Date | R(AU) | Δ(AU) | PA(°)[a] | Time$_L$ (min)[b] | Site | Instrument | Comments |
|---|---|---|---|---|---|---|---|
| 2001-12-08 | 31.386 | 31.147 | 1.749 | 259.05 | HST | STIS | not tracked |
| 2001-12-09 | 31.385 | 31.316 | 1.756 | 259.18 | HST | STIS | not tracked |
| 2003-06-30 | 31.205 | 31.165 | 1.868 | 259.19 | HST | ACS | — |
| 2003-07-09 | 31.203 | 31.013 | 1.842 | 257.93 | HST | ACS | — |
| 2003-07-25 | 31.198 | 30.756 | 1.695 | 255.79 | HST | ACS | — |
| 2003-08-23 | 31.190 | 30.380 | 1.134 | 252.66 | HST | ACS | — |
| 2004-05-26 | 31.111 | 31.661 | 1.551 | 263.32 | HST | ACS | — |
| 2005-08-12 | 30.996 | 30.363 | 1.482 | 252.52 | HST | ACS | — |
| 2006-07-07 | 30.919 | 30.880 | 1.885 | 256.82 | HST | ACS | — |
| 2008-09-26 | 30.763 | 29.800 | 0.531 | 247.84 | Bigelow | Mont4k | Photometric |
| 2008-09-27 | 30.763 | 29.796 | 0.504 | 247.81 | Bigelow | Mont4k | Photometric |
| 2008-09-30 | 30.763 | 29.787 | 0.428 | 247.73 | CTIO | Mosaic | Photometric |
| 2008-10-01 | 30.762 | 29.784 | 0.404 | 247.70 | CTIO | Mosaic | Photometric |
| 2008-10-02 | 30.762 | 29.782 | 0.381 | 247.68 | CTIO | Mosaic | Cirrus |

[a] solar phase angle
[b] light travel time

TABLE 2. ASTROMETRY

| Date | Hour | RA[a] | Dec[b] | ΔRA_B[c] | ΔDec_B[c] | ΔRA_A2[c] | ΔDec_A2[c] |
|---|---|---|---|---|---|---|---|
| 2001/12/08 | 21.71786 | 00:10:33.41 | -08:02:43.9 | 0.0019±0.0088 | 0.3812±0.0084 | -0.0247±0.0182 | 0.0367±0.0124 |
| 2001/12/09 | 07.34166 | 00:10:32.99 | -08:02:32.5 | 0.0385±0.0114 | 0.3750±0.0082 | 0.0324±0.0113 | 0.0244±0.0218 |
| 2003/06/30 | 01.70880 | 00:36:49.57 | -05:11:28.2 | 0.2715±0.0009 | 0.1839±0.0005 | -0.0238±0.0008 | -0.0084±0.0010 |
| 2003/07/09 | 19.41372 | 00:37:00.11 | -05:13:14.5 | 0.2278±0.0003 | -0.1081±0.0007 | -0.0210±0.0007 | 0.0154±0.0008 |
| 2003/07/25 | 17.93292 | 00:36:53.80 | -05:18:29.9 | -0.1576±0.0014 | -0.0296±0.0003 | 0.0236±0.0013 | 0.0040±0.0017 |
| 2003/08/23 | 13.35984 | 00:35:33.71 | -05:33:53.5 | 0.3146±0.0015 | 0.0447±0.0014 | 0.0228±0.0025 | -0.0214±0.0021 |
| 2004/05/26 | 18.32280 | 00:43:06.70 | -04:21:34.9 | -0.1431±0.0005 | 0.1474±0.0013 | -0.0192±0.0018 | -0.0241±0.0012 |
| 2005/08/12 | 23.45220 | 00:53:20.19 | -03:36:42.1 | -0.0249±0.0031 | -0.2734±0.0021 | 0.0223±0.0028 | 0.0167±0.0028 |
| 2006/07/06 | 12.96780 | 01:02:14.64 | -02:27:58.6 | 0.2583±0.0019 | 0.2702±0.0019 | -0.0174±0.0011 | 0.0258±0.0004 |

[a] HH:MM:SS.SS
[b] DD:MM:SS.S
[c] arcseconds

TABLE 3. A1,A2 MODEL FITTING RESULTS

| Rootname | Julian date | $\Delta x^a$ | $\Delta y^a$ | f2/f1 | $\chi_v^2$ A | $\chi_v^2$ A1,A2 |
|---|---|---|---|---|---|---|
| j8rl05abq | 2452820.55979 | 0.162 | -1.031 | 0.86 | 4.79 | 2.76 |
| j8rl05acq | 2452820.56740 | 0.174 | -1.025 | 0.83 | 3.07 | 1.01 |
| j8rl05adq | 2452820.57500 | 0.184 | -0.949 | 0.76 | 2.68 | 1.31 |
| j8rl05aeq | 2452820.58260 | 0.227 | -1.003 | 0.65 | 2.88 | 1.20 |
| j8rl06x0q | 2452830.29750 | 0.720 | -0.749 | 1.07 | 3.05 | 0.94 |
| j8rl06x1q | 2452830.30510 | 0.743 | -0.714 | 0.97 | 3.15 | 1.31 |
| j8rl06x2q | 2452830.31271 | 0.749 | -0.685 | 0.92 | 3.48 | 1.99 |
| j8rl06x3q | 2452830.32031 | 0.768 | -0.668 | 0.93 | 3.13 | 1.63 |
| j8rl07h8q | 2452846.23580 | 0.151 | 0.907 | 1.02 | 3.04 | 1.62 |
| j8rl07h9q | 2452846.24340 | 0.083 | 0.991 | 0.89 | 3.06 | 1.79 |
| j8rl07haq | 2452846.25101 | 0.031 | 1.013 | 0.87 | 2.48 | 1.23 |
| j8rl07hbq | 2452846.25861 | 0.012 | 0.914 | 1.08 | 2.29 | 1.44 |
| j8rl08b9q[b] | 2452875.04526 | -0.785 | 0.718 | 0.72 | 5.76 | 3.86 |
| j8rl08baq | 2452875.05286 | -0.864 | 0.822 | 0.65 | 4.60 | 1.33 |
| j8rl08bbq | 2452875.06046 | -0.882 | 0.912 | 0.75 | 4.34 | 1.10 |
| j8rl08bcq | 2452875.06807 | -0.992 | 0.926 | 0.71 | 5.14 | 1.24 |
| j8rl09p9q | 2453152.25204 | -0.508 | -1.057 | 1.23 | 4.35 | 1.17 |
| j8rl09paq | 2453152.25965 | -0.458 | -1.071 | 1.03 | 3.97 | 1.47 |
| j8rl09pbq | 2453152.26725 | -0.386 | -1.103 | 1.05 | 3.57 | 1.22 |
| j8rl09pcq | 2453152.27485 | -0.369 | -1.176 | 1.25 | 4.05 | 1.46 |
| j8rl19akq | 2453595.46545 | 0.682 | 0.826 | 0.87 | 4.08 | 1.06 |
| j8rl110alq[b] | 2453595.47306 | 0.707 | 0.953 | 2.30 | 5.83 | 3.56 |
| j8rl110amq | 2453595.48129 | 0.547 | 0.720 | 0.90 | 2.10 | 1.36 |
| j8rl110anq | 2453595.48889 | 0.501 | 0.784 | 1.55 | 2.21 | 1.34 |
| j8rl04r8q | 2453923.02892 | 1.066 | -0.381 | 1.34 | 4.72 | 1.86 |
| j8rl04r9q | 2453923.03652 | 1.107 | -0.412 | 1.03 | 5.13 | 1.69 |
| j8rl04raq | 2453923.04413 | 1.075 | -0.361 | 1.01 | 4.44 | 1.53 |
| j8rl04rbq | 2453923.05173 | 1.050 | -0.316 | 1.07 | 3.99 | 1.61 |

[a] measured in pixels
[b] cosmic ray poorly placed, fitting difficult

TABLE 4. ORBITAL ELEMENTS AND DERIVED PARAMETERS

| Parameter | | A2 Solution[c,d] | B Solution[d] |
|---|---|---|---|
| Fitted orbital elements | | | |
| Period (days) | P | 1.9068±0.0001 | 50.302±0.001 |
| Semi-major axis (km) | a | 867±11 | 7411±12 |
| Eccentricity | e | 0.101±0.006 | 0.2949±0.0009 |
| Inclination (°) | i | 88.9±0.6 | 79.3±0.2 |
| Mean longitude[a] at epoch[b] (°) | $\varepsilon$ | 184.4±1.6 | 281.1±0.3 |
| Longitude of Ascending Node[a] (°) | $\Omega$ | 330.0±1.0 | 325.2±0.1 |
| Longitude of periapsis[a] (°) | $\varpi$ | 47.7±6.3 | 292.1±0.2 |
| Derived Parameters | | | |
| System Mass (x$10^{18}$ kg) | $M_{sys}$ | 14.20±0.05 | 12.75±0.06 |
| Orbit pole right ascension[a] (°) | $\alpha$ | 240.055±1.149 | 236.837±0.162 |
| Orbit pole declination[a] (°) | $\delta$ | 1.023±0.389 | 11.167±0.175 |

[a] Referenced to J2000 equatorial frame.
[b] The epoch is Julian date 2453880 (2006 May 24 12:00 UT).
[c] excludes the STIS observations as discussed in Section 4.2
[d] Best fits to the data. The chi-square values are relatively large because precession is not included in our model and the data span a long enough timespan that precession is observed.

TABLE 5. MULTIPLE SYSTEMS

| System | Region[a] | $a_\odot$ (AU) | $M_{sys}$ (kg) | $\rho$ (kg m$^{-3}$) | $a_1$ (km) | $a_2$ (km) | $d_0$ (km) | $d_1$ (km) | $d_2$ (km) | $P_1$ (days) | $P_2$ (days) | $T_o$ (hr) | Ref |
|---|---|---|---|---|---|---|---|---|---|---|---|---|---|
| (136617) 1994 CC | NEA | 1.64 | 1.4x10$^{11}$ | ***1000*** | ***0.5*** | ***1.2*** | ***0.65*** | ***0.05*** | ***0.1*** | ***0.2*** | ***0.83*** | 2.39 | (1) |
| (153591) 2001SN$_{263}$ | NEA | 1.98 | 1.2x10$^{13}$ | 1000 | 4 | 17 | 2.8 | 0.5 | 1.2 | 1.92 | 6.13 | 3.4 | (2) |
| (3749) Balam | MBA | 2.23 | 1.3x10$^{14}$ | 1200 | 20 | 310 | 7 | ***3*** | 1.5 | 1.39 | 110 | 2.80 | (3) |
| (45) Eugenia | MBA | 2.72 | 5.8x10$^{18}$ | 1120 | 700 | 1184 | 214.6 | 6 | ***12.7*** | ***2*** | 4.76 | 5.7 | (4) |
| (216) Kleopatra | MBA | 2.79 | 2.3x10$^{18}$ | 3500 | 380 | 650 | 118 | 3 | 5 | ***1.4*** | 4.2 | 5.39 | (5) |
| (87) Sylvia | MBA | 3.49 | 1.5x10$^{19}$ | 1200 | 706 | 1356 | 287 | 7 | 18 | 1.38 | 3.64 | 5.18 | (6) |
| (47171) 1999 TC$_{36}$ | KB | 39.2 | 1.3x10$^{19}$ | 582 | 867 | 7411 | 281 | 259 | 135 | 1.90 | 50.3 | 45.7[c] | (7) |
| Pluto[b] | KB | 39.5 | 1.3x10$^{22}$ | 1920 | 19571 | 49240 | 2412 | 1212 | 88 | 6.38 | 25.49 | 153.3 | (8) |
| Haumea | KB | 43.1 | 4.4x10$^{21}$ | 3000 | 25657 | 49880 | 1436 | 160 | 320 | 34.7 | 49.1 | 3.91 | (9) |

Note: In the column headings, subscripts 0, 1 and 2 denote the primary, 1st component, and 2$^{nd}$ component, respectively. $a$ is semi-major axis, $d$ is the diameter of the component, $P$ is the orbital period and $T$ is the rotational period of the primary.
[a] NEA=Near Earth Asteroids, MBA = Main Belt Asteroids, KB = Kuiper Belt
[b] fourth component $a_3$=65210 km, $d_3$=72 km, $P_3$=38.85 days.
[c] assumed to be synchronous with orbital period.
References: (1) IAUC 9053; (2) Nolan et al. 2008, Betzler et al. 2008; (3) IAUC 8928, IAUC 7827, CBET 1297; (4) IAUC 8817 (5) IAUC 8980; (6) Marchis et al. 2005; (7) This work; (8) Tholen et al. 2008; (9) Ragozzine et al. 2009; Rabinowitz et al. 2006.
Values in bold/italics are come from the Johnston "Asteroids with Satellites" archive (http://www.johnstonsarchive.net/astro/asteroidmoons.html). System mass is calculated from the values provided in the table.

TABLE 6. PHOTOMETRY

| Julian Date | VisitID | Filter | $N_{OBS}$ | $M_{TOTAL}$ | $\Delta m_{A1A2}$ | $\Delta m_{AB}$ |
|---|---|---|---|---|---|---|
| 2452820.57120 | 05 | F606W | 4 | 19.571±0.009 | 0.28±0.07 | 2.17±0.02 |
| 2452830.30890 | 06 | F606W | 4 | 19.546±0.013 | 0.0±0.0 | 2.16±0.04 |
| 2452846.24720 | 07 | F606W | 4 | 19.520±0.011 | 0.0±0.0 | 2.20±0.03 |
| 2452875.05666 | 08 | F606W | 4 | 19.409±0.007 | 0.37±0.03 | 2.22±0.01 |
| 2453152.26345 | 09 | F606W | 4 | 19.556±0.014 | 0.14±0.05 | 2.24±0.03 |
| 2453595.46925 | 10 | F606W | 2 | 19.429±0.014 | 0.3±0.3 | 2.27±0.03 |
| 2453595.48509 | 10 | F814W | 2 | 18.388±0.014 | 0.2±0.2 | 2.30±0.03 |
| 2453923.04032 | 04 | F606W | 4 | 19.488±0.002 | 0.11±0.07 | 2.36±0.12 |